\documentclass[aps,prl,twocolumn,superscriptaddress,showpacs]{revtex4}

\usepackage{graphicx,amsmath,amssymb,times}
\newcommand{\s}{\sigma}

\newcommand{\pSG}{p_{\rm{SG}}}
\newcommand{\TSG}{T_{\rm{SG}}}
\newcommand{\eqd}{\stackrel{\text{d}}{=}}
\newcommand{\sign}{\text{sign}}

\begin{document}

\title{Entropic Effects in the Very Low Temperature Regime of Diluted
  Ising Spin Glasses\\ with Discrete Couplings}

\author{Thomas J\"org} \affiliation{LPTMS, Universit\'e de Paris-Sud,
  B\^atiment 100, 91405 Orsay Cedex, France} \affiliation{\'Equipe TAO---INRIA 
  Futurs, 91405 Orsay Cedex, France}

\author{Federico Ricci-Tersenghi} \affiliation{Dipartimento di Fisica
  and INFM-CNR, Universit\`{a} di Roma ``La Sapienza'', P.le Aldo Moro
  2, 00185 Roma, Italy}

\date{\today}

\begin{abstract}
  We study link-diluted $\pm J$ Ising spin-glass models on the
  hierarchical lattice and on a three-dimensional lattice close to the
  percolation threshold.  We show that previously computed zero
  temperature fixed points are unstable with respect to temperature
  perturbations and do not belong to any critical line in the
  dilution-temperature plane.  We discuss implications of the presence
  of such spurious unstable fixed points on the use of optimization
  algorithms, and we show how entropic effects should be taken into
  account to obtain the right physical behavior and critical points.
\end{abstract}

\pacs{75.10.Nr, 64.60.Ak, 75.40.Mg, 75.50.Lk}

\maketitle

\paragraph*{Introduction.---}\hspace{-0.45cm}
Frustrated systems may have very complex free-energy landscapes at low
temperature $T$ which in turn may give rise to very peculiar
thermodynamical properties
\cite{villain:80,parisi:79,mezard:87,diep:05}. It was recently shown
in Ref.~\cite{joerg:06a} for the case of the two-dimensional
Edwards-Anderson (EA) model that in such situations $T=0$ computations
as the ones of Ref.~\cite{amoruso:03} may produce misleading
results. In Ref.~\cite{hartmann:07} it was shown that there are ways
to improve on the results of Ref.~\cite{amoruso:03} using $T=0$
methods; however the $\eta$ exponent is still found to be
non-universal; this could be due to entropic effects, which are
neglected in Ref.~\cite{hartmann:07}.  In this Letter we study
entropic effects in the very low temperature regime of disordered
models.  We consider diluted spin glasses (SG) with discrete coupling
distributions at very low temperature and show that previous
computations of critical points were incorrect
\cite{bray:87b,boettcher:04b}. We explain why and we show how to
compute the right critical points: considering first-order corrections
in $T$ allows us to treat the entropic contribution to the free energy
correctly in the limit when $T$ goes to zero.  A similar idea has been
applied also to models defined on random graphs \cite{PNAS}.  Here we
find that, on hierarchical as well as three-dimensional (3D) lattices, the SG phase
persists in a region of higher dilution than calculations done exactly
at $T=0$ would predict. In other words, we find SG ordering induced
solely by entropic effects which is a phenomenon somewhat reminiscent
of Villain's ``order by disorder'' \cite{villain:80,note_gapless}.

We consider two SG models: one defined on the hierarchical lattice
\cite{berker:79}, where the Migdal-Kadanoff approximation is
correct, and the other one on a 3D lattice. In the
former case, we are able to compute the distribution of effective
couplings and of the free energies allowing us to determine exactly
the location of the critical points and the whole phase diagram. For
the 3D case, we have to resort to numerical simulations: using
state-of-the-art Monte Carlo (MC) methods for disordered systems, we
provide evidence that previously computed critical lines are likely to
be incorrect, and by a numerical study of percolation properties we
give bounds on the location of the right critical points.
\begin{figure}
  \includegraphics[width=\columnwidth]{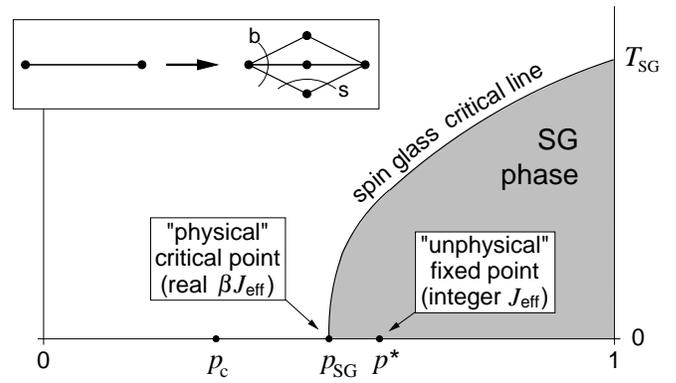}
  \caption{Phase diagram of the diluted $\pm J$ spin glass in the
    temperature versus bond density plane, where $p_{\rm{c}}$ denotes the
    geometric percolation threshold, $\pSG$ the percolation threshold
    for spin-glass order and $p^*$ the percolation threshold for spin-glass 
    order determined at $T=0$. Inset: Elementary step for the
    construction of a hierarchical lattice with $b\!=\!3$ and
    $s\!=\!2$.}
  \label{phaseDiagram}
\end{figure}

\paragraph*{Model.---}\hspace{-0.45cm}
We consider an Ising SG model defined by the Hamiltonian
\begin{equation}
  H(\underline\s) = \sum_{(i,j)} J_{ij} \s_i \s_{\!j}\!,
  \label{ham}
\end{equation}
with Ising spins $\s_i = \pm1$ and the sum running over all the
nearest neighbors pairs on the lattice. The couplings $J_{ij}$ are
quenched, independent and identically distributed random variables
extracted from the distribution \cite{shapira:94}
\begin{equation}
  P_0(J) = (1-p) \delta(J) + \frac{p}{2} \Big[\delta(J - 1) + \delta(J +
  1)\Big]\!.
  \label{P0}
\end{equation}
We are interested in the $T=0$ critical point separating the
paramagnetic phase ($p<\pSG$) from the SG phase ($p>\pSG$). We
consider two lattices: the hierarchical lattice and the 3D simple
cubic lattice for which the Hamiltonian in Eq.~\eqref{ham} corresponds
to the EA model \cite{edwards:75}.

\paragraph*{Hierarchical lattice.---}\hspace{-0.45cm}
The hierarchical lattice of $G$ generations is obtained by applying
the construction shown in the inset of Fig.~\ref{phaseDiagram} to all
the links of the $G-1$ generations lattice (the $G=0$ lattice being a
single link connecting two vertices). Such a construction is defined
by two parameters: the number $b$ of parallel branches, made of $s$
bonds in series each.  The effective dimension of the model is defined
by $d=1+\ln(b)/\ln(s)$ and it is known that the lower critical
dimension for the SG transition is close to $d_\ell \simeq 2.5$
\cite{southern:77,nifle:92,amoruso:03}. Working with $d > d_\ell$ we
have $T_\text{SG}>0$ for $p=1$ and a critical line in the $(p,T)$
plane (see Fig.~\ref{phaseDiagram}).  We present results for $b=3$ and
$s=2$ (i.e., $d=2.585...$), but we have checked their validity also
for other choices of the parameters.

A model defined on the hierarchical lattice can be solved exactly by
recursive decimation: starting from a lattice of $G$ generations and
summing over the spins introduced in the last generation one can
easily obtain a lattice of $G-1$ generations with renormalized
couplings.  At each step of this renormalization procedure the
couplings undergo two elementary transformations: (i) each $s$-tuple
of couplings in series produces an effective coupling $\tilde{J}$ and
(ii) each $b$-tuple of effective couplings in parallel linking the
same pair of variables is summed together, giving the renormalized
coupling.  In the $T \to 0$ limit, one renormalization step can be
written as
\begin{align}
  &\tilde{J}^{(n)} \eqd \sign(J^{(n)}_1 \cdot \ldots \cdot J^{(n)}_s)
  \min(|J^{(n)}_1|,\ldots,|J^{(n)}_s|)\!,\label{rec1}\\
  &J^{(n+1)} \eqd \tilde{J}^{(n)}_1 + \ldots +
  \tilde{J}^{(n)}_b\!,\label{rec2}
\end{align}
where equalities hold in distribution sense and the index $(n)$ stays
for the number of renormalization steps.  Applying these functional
recursion equations, the initial distribution $P_0$ of couplings flows
to the fixed-point one $P_\infty$.  The critical density of links
$p^*$ is defined such that $P^*_\infty$ is nontrivial, i.e.,
different from the paramagnetic one, $\delta(J)$, and the SG one,
$[\delta(J-\infty) + \delta(J+\infty)]/2$.

\begin{figure}
  \includegraphics[width=\columnwidth]{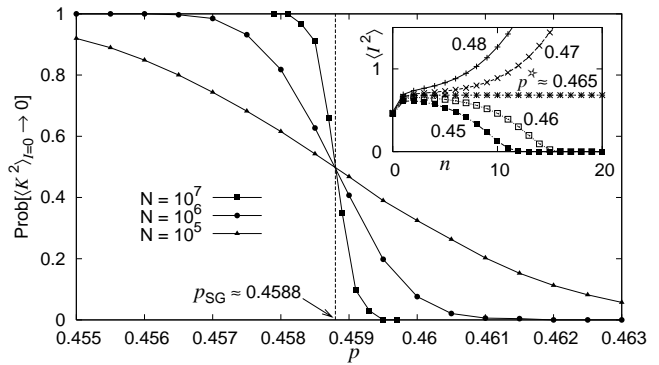}
  \caption{The probability that the variance of the renormalized
    ``entropic'' couplings $K$ goes to zero as a function of bond
    density $p$, for different sizes $N$ of the population evolved
    numerically.  The inset shows the behavior of the variance of the
    renormalized ``energetic'' couplings $I$, obtained analytically in
    the $N \to\infty$ limit.}
  \label{2fields}
\end{figure}

In Ref.~\cite{bray:87b} Bray and Feng computed the distribution
$P_\infty^*$ under the assumption that the renormalized couplings
remain integer, which is indeed a possible solution to
Eqs.~\eqref{rec1} and \eqref{rec2}.  Actually in Ref.~\cite{bray:87b}
only three values for the couplings (0, $\pm 1$) were considered; the
extension to a symmetric distribution involving all the integer values
between $-M$ and $M$ is straightforward, and leads in the limit of
large $M$ to $p^*\simeq0.465$.  In the inset of Fig.~\ref{2fields} we
show the variance of $J^{(n)}$ as a function of $n$ for various values
of $p$.  This critical point $p^*$ has been considered up to now the
boundary between the paramagnetic and the SG phases.  Nevertheless, we
find that in the plane $(p,T)$ the point $(p^*,0)$ does not belong to
{\it any} critical line; i.e., it is an isolated point (see
Fig.~\ref{phaseDiagram}), and for this reason it is irrelevant for the
thermodynamics at any positive $T$. Even at $T=0$ it does not
correspond to the border between two different phases, i.e., $p^* \neq
\pSG$!

In order to prove that the fixed point $(p^*,0)$ is isolated as in
Fig.~\ref{phaseDiagram} we study coupling renormalization at
infinitesimal $T$. We start from the positive $T$ rule for decimating
$s=2$ bonds in series (for $s>2$ the same transformation must be
applied more than once)
\begin{equation}
  \tanh(\beta \tilde{J}) = \tanh(\beta J_1) \tanh(\beta J_2)\!,
  \label{finiteT}
\end{equation}
with $\beta = 1/T$.  Similarly to \cite{parisi:05}, for very low $T$,
we rewrite $J = T K$ in case $J$ is vanishing for $T \to 0$, and $J =
\sign(I) (|I| - T K)$ if $\lim_{T \to 0} J = I \neq 0$.  The choice in
the sign is dictated by the fact that thermal fluctuations decrease
the coupling intensity.  We have checked that higher order corrections
in temperature are unnecessary in the $T \to 0$ limit.  Plugging these
new variables in Eq.~\eqref{finiteT} we find that in the $T \to 0$
limit the ``energetic'' component $I$ gets renormalized following
Eq.~\eqref{rec1}, while the ``entropic'' component $K$ follows
\begin{align}
  \tilde{K} &= K_1                     &&
  \text{if  } |I_1| < |I_2| \label{eqIneq} \\
  \tanh(\tilde{K}) &= \tanh(K_1) \tanh(K_2) &&
  \text{if } I_1 = I_2 = 0 \label{eqI0}\\
  \exp(2\tilde{K}) &= \exp(2 K_1) + \exp(2 K_2) &&
  \text{if } I_1 = I_2 \neq 0. \label{eqInot0}
\end{align}
Before renormalization, the original couplings are
$T$-in\-de\-pen\-dent and distributed according to $P_0$; thus we
start with $I \in \{0,\pm 1\}$ and $K=0$.  The recursive equation for
$I$ is the one already studied above and gives a coupling flowing to 0
for $p < p^*$ (see inset of Fig.~\ref{2fields}).  Nevertheless, even
if $I=0$ for all the renormalized couplings, the entropic correction
may be non-null, thus giving a nonzero correlation of the order of
$\tanh(\beta J) = \tanh(K) \neq 0$, even in the $T \to 0$ limit.
Please note that if the temperature was set to zero at the beginning
of the calculation such a nonzero correlation would not be found.

Under renormalization the variance of entropic couplings goes to zero
or diverges, depending on the value of $p$.  In the main panel of
Fig.~\ref{2fields} we show the probability that $\langle K^2
\rangle_{I=0}$ flows to zero, where $N$ is the size of the population
used for simulating the decimation procedure.  It clearly shows the
existence of a critical value $\pSG \simeq 0.4588$ such that for $p
< \pSG$ correlations decay to zero at large distances, while for $p >
\pSG$ there are long-range correlations (of SG type, since $\langle K
\rangle = 0$).

On hierarchical lattices, the inequality $\pSG < p^*$ holds in
general, and implies that for $T=0$ and $p \in (\pSG, p^*)$ the long
range-order is induced solely by entropic effects. That is the two
ground states obtained by fixing the outermost spins of the
hierarchical lattice in a parallel or anti-parallel way have exactly
the same energy, but very different entropies; long-range correlations
from the ground state with the largest entropy will dominate ensemble
averaged correlations.  Such long-range correlations arise because
when summing two parallel effective bonds such that $\tilde{I}_1 +
\tilde{I}_2 = 0$ it may be that $\tilde{K}_1 + \tilde{K}_2 \neq 0$,
i.e., there is no perfect cancellation and an effective coupling of
intensity $\mathcal{O}(T)$ persists.  In the $T \to 0$ limit these
couplings induce nonzero correlations, which may eventually increase
under renormalization (as for $p > \pSG$).  It is worth noticing that
this mechanism for generating $\mathcal{O}(T)$ couplings may work
perfectly well also when applying a decimation procedure on regular
lattices \cite{boettcher:04b}.

A similar argument also tells us that on the very first steps of
decimation, when $K$ takes few and discrete values (for large $n$, $K$
becomes dense on $\mathbb{R}$), exact cancellations, both in $I$ and
$K$, take place.  If the original model had a bond density slightly
larger than the percolating density $p_{\rm c}$ (for $b=3$ and $s=2$,
$p_{\rm c}=0.389391...$) these exact cancellations would produce a renormalized
lattice which is below the percolating point, preventing any long-range 
correlation from arising.  So also the inequality $p_{\rm c} < \pSG$ must
hold in general for any frustrated model with discrete couplings on
the hierarchical lattice.

Regarding the issue of the universality, we checked that starting from
any point on the critical line extending from $(\pSG,0)$ to
$(1,\TSG)$, see Fig.~\ref{phaseDiagram}, the same fixed-point coupling
distribution is obtained; note that, at a positive temperature,
couplings $J/T$ get renormalized exactly by Eq.~\eqref{eqI0}.  This
fixed-point distribution has nothing to do with the integer-valued one
obtained from $(p^*,0)$.

\paragraph*{3D simple cubic lattice.---}\hspace{-0.45cm}
The phase diagram shown in Fig.~\ref{phaseDiagram} is exact for the SG
model defined by Eq.~\eqref{ham} with coupling distribution from
Eq.~\eqref{P0} on the hierarchical lattice (or equivalently in the Migdal-Kadanoff
approximation). We present now evidence that the main qualitative
features of that phase diagram are preserved when a 3D lattice is
considered. With this in mind, we show by MC simulations and by percolation
arguments that (i) a SG transition takes place at finite temperature
for $p < p^*$, thus implying $\pSG < p^*$ and that (ii) there exists a
percolating phase with no long-range order, thus implying $p_{\rm c} < \pSG$
with $p_{\rm c} = 0.2488126(5)$ \cite{lorenz:98}.

The MC technique we use to simulate the 3D link-diluted $\pm J$ EA
model close to the percolation threshold is a combination of the
replica cluster update moves \cite{clusterAlgo} embedded in parallel
tempering \cite{hukushima:96} as described in Ref.~\cite{joerg:06} in
alternation with standard Swendsen-Wang \cite{swendsen:87}
cluster updates on each of the replicas. The additional Swendsen-Wang updates are
very efficient close to percolation and speed up the simulation
considerably \cite{persky:96}. This is an important issue because close to
percolation we encounter noticeable finite-size effects, and therefore
relatively large lattices have to be simulated to see a good signal
for a SG transition below $p^*$. We use the value $p^*=0.272(1)$
determined by Boettcher \cite{boettcher:04b} from the behavior of the
domain-wall defect energies of the ground states using an optimization
algorithm \cite{boettcher:01}.

To check for the presence of a SG transition we measure the
second-moment correlation length $\xi(L,T)$ on a lattice of size $L$
defined as \cite{corrLength}
\begin{equation}
  \label{eq:correlation_length}
  \xi(L,T) = \frac{1}{2 \sin(|{\mathbf k}_{\rm min}|/2)} 
  \left[ \frac{\chi_{\rm SG}(\mathbf 0)}{\chi_{\rm SG}({\mathbf k}_{\rm min})} - 1
  \right]^{1/2} \!,
\end{equation}
where the wave-vector-dependent SG susceptibility is
\begin{equation}
  \label{eq:susceptibility}
  \chi_{\rm SG}({\mathbf k}) = \frac{1}{L^3} \sum_{\mathbf x}\sum_{\mathbf
  r} e^{\imath {\mathbf k \mathbf r}} \overline{\langle q_{\mathbf x}
  q_{\mathbf x + \mathbf r} \rangle}
\end{equation}
and ${\mathbf k}_{\rm min} = (0,0,2 \pi/L)$ is the smallest nonzero
wave vector allowed by periodic boundary conditions.  We denote thermal
averages by $\langle \cdot \rangle$ and disorder averages by
$\overline{\,\cdot\,}$.

\begin{figure}
  \includegraphics[width=0.49\columnwidth]{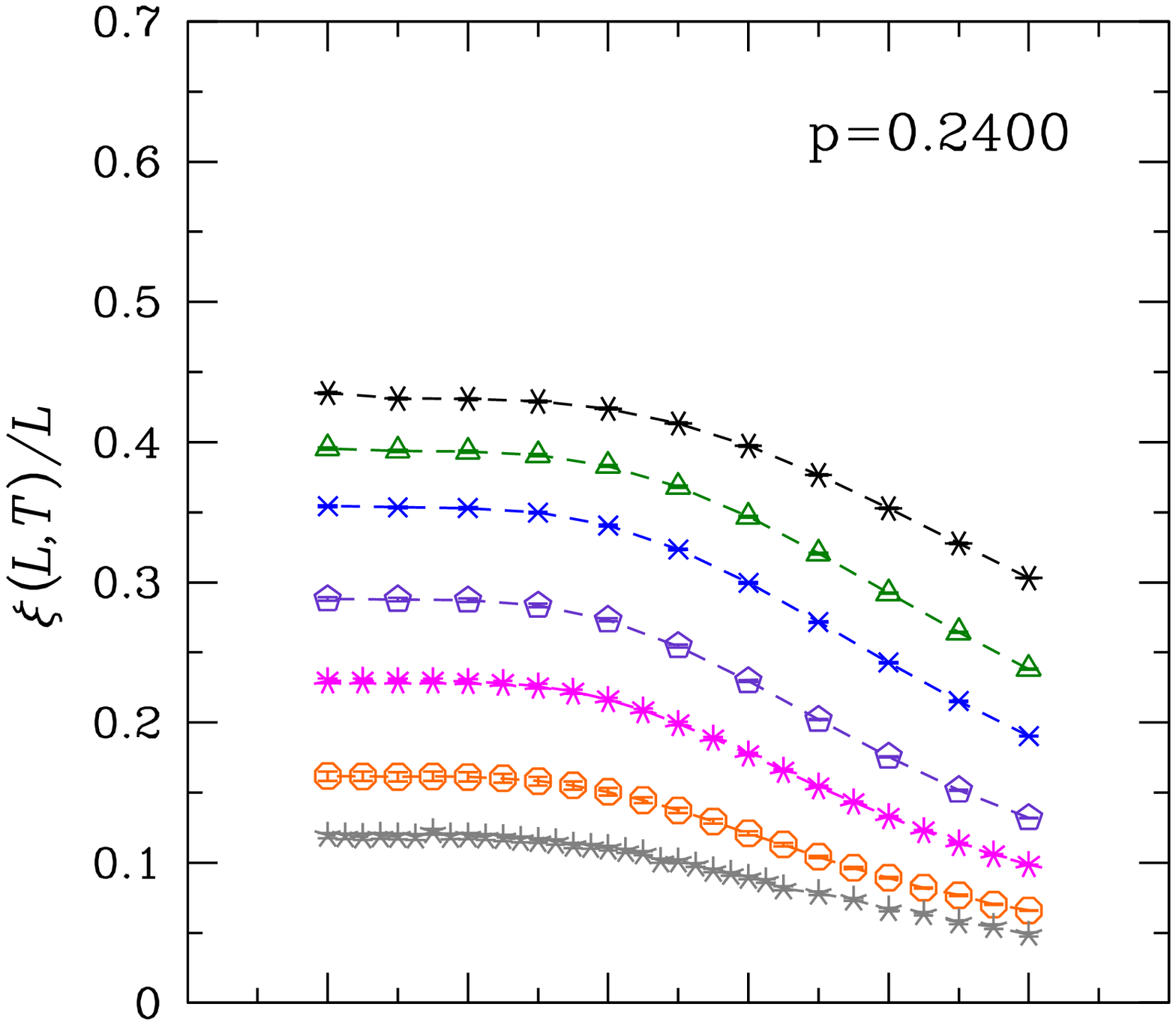}\hspace*{-0.7cm}
  \includegraphics[width=0.49\columnwidth]{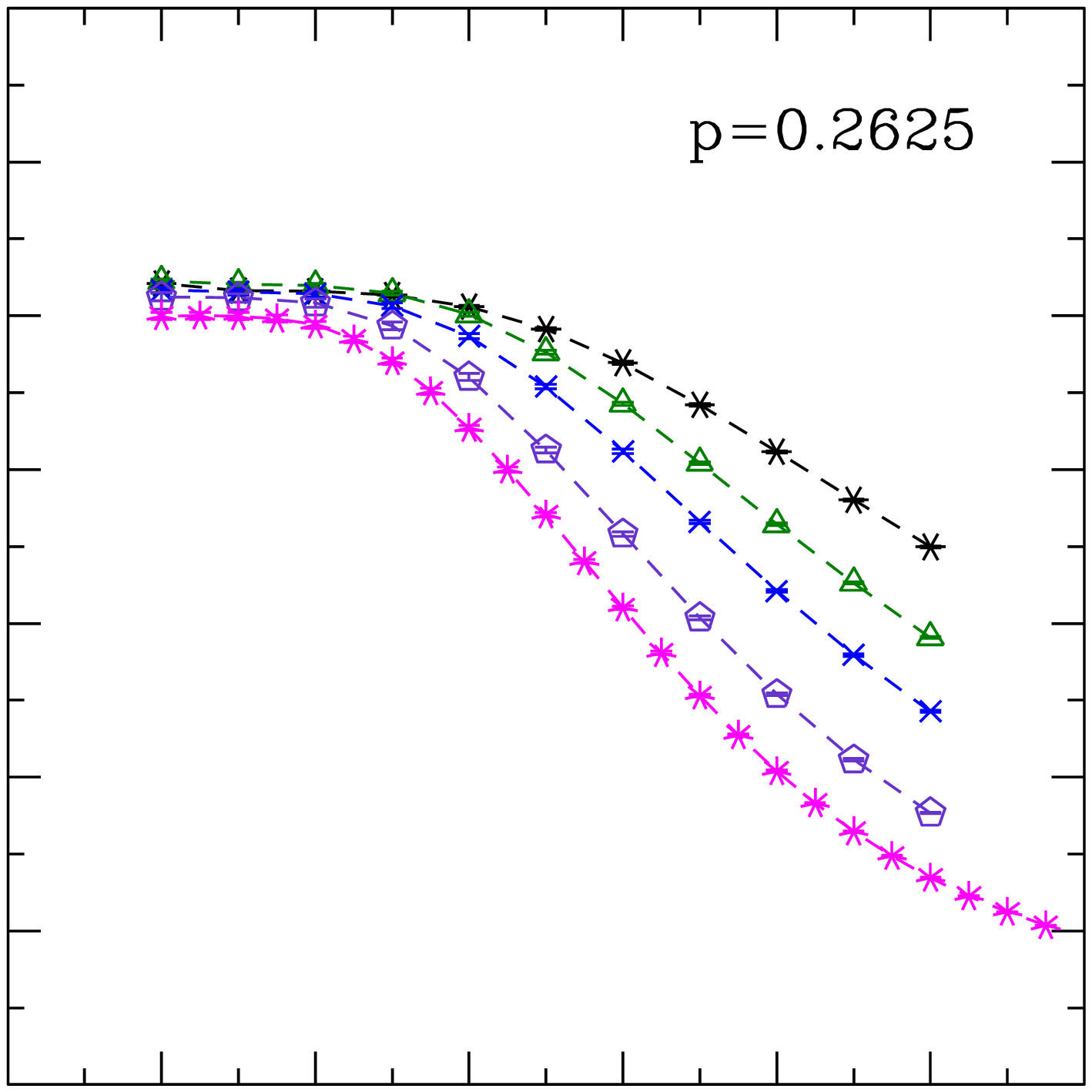}\vspace*{-0.5cm}
  \includegraphics[width=0.49\columnwidth]{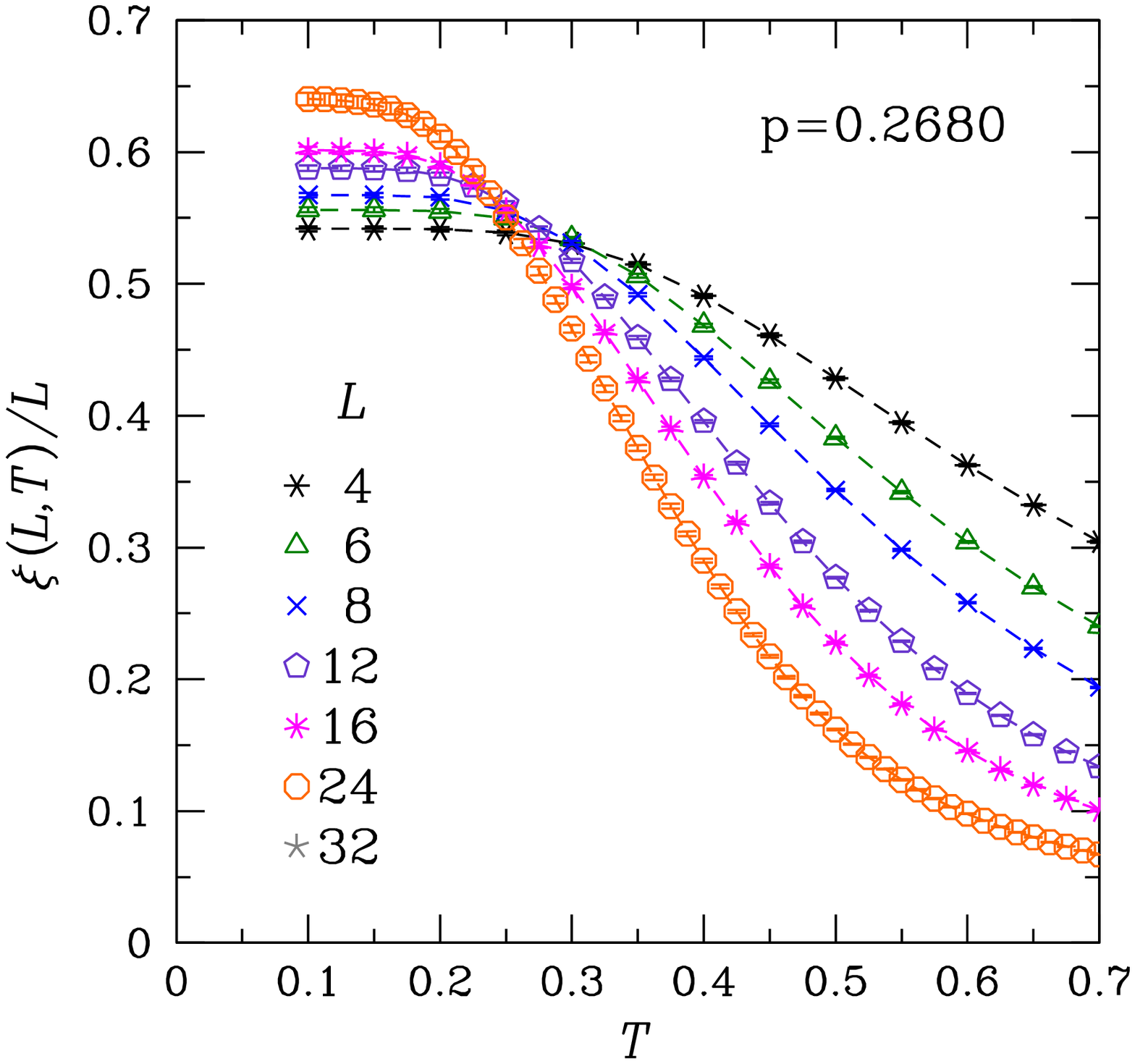}\hspace*{-0.7cm}
  \includegraphics[width=0.49\columnwidth]{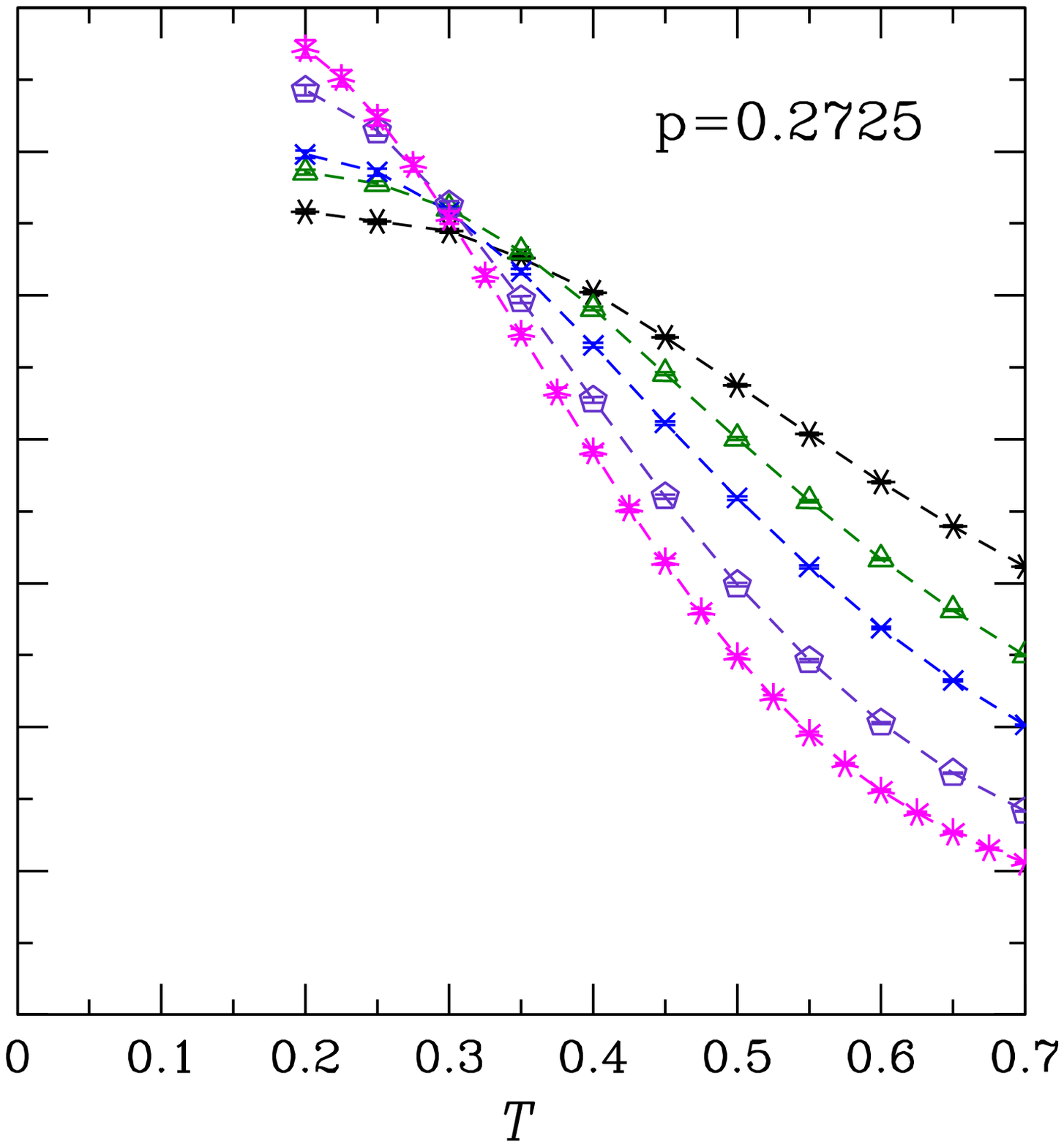}
  \caption{(Color online). $\xi(L,T)/L$ as a function of temperature 
    for several system sizes $L$ and different bond densities $p$.}
  \label{fig:xi_from_MC}
  \vspace*{-0.2cm}
\end{figure}
In Fig.~\ref{fig:xi_from_MC} we display the behavior of $\xi(L,T)/L$
for four different values of $p$. In the upper left-hand panel we show the
result below the percolation threshold ($p = 0.24 < p_{\rm c}$) for which
there is no SG transition as can be seen from the absence of a
crossing of $\xi(L,T)/L$ for different lattice sizes. The upper right-hand
panel shows the situation for $p_{\rm c} < p = 0.2625 < p^*$ where again the
curves do not cross and therefore there is no sign for a SG
transition. For $p = 0.268$, slightly below $p^*$, we find --- amongst
noticeable finite-size effects --- that the $\xi(L,T)/L$ for different
lattice sizes do cross indicating the presence of a SG transition.
The crossing point is moving slightly towards a smaller effective
$T_c$ as $L$ is increased, but this is also the case for $p = 0.2725
\sim p^*$ and larger values of $p$.  The fact that the crossing
happens at a value of $\xi(L,T_c)/L \approx 0.6$, with a tendency to
increase for larger $L$, is an important evidence for a SG transition
since this value is a renormalization group invariant quantity and was
found to lie between $0.60$ and $0.65$ in different recent studies of
the 3D SG transition \cite{ballesteros:00,joerg:06,katzgraber:06}.

\begin{figure}
  \includegraphics[width=\columnwidth]{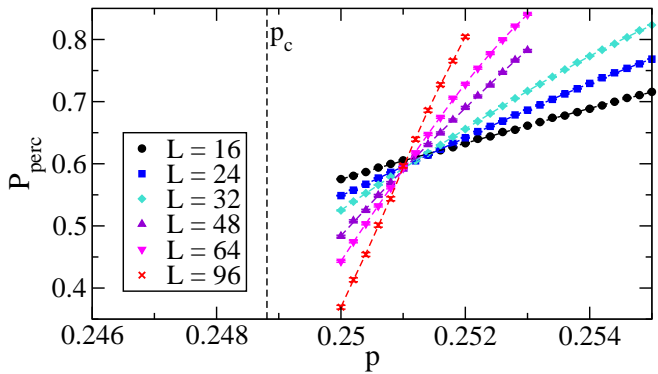}
  \caption{(Color online). Percolation probability in a 3D $\pm J$
    diluted spin-glass model, reduced with $T=0^+$ rules.}
  \label{fig:percolation_at_T}
  \vspace*{-0.2cm}
\end{figure}

To show that indeed $p_{\rm c}\!<\!\pSG$ holds in 3D, we apply to the
Hamiltonian in Eq.~\eqref{ham} the reduction rules for weakly
connected lattice sites (i.e., with degree 1 or 2) as described in
Ref.~\cite{boettcher:04b}.  Instead of applying these rules exactly at
$T\!=\!0$ we again keep terms of order $T$ for the resulting effective
couplings leading to a more connected remaining graph than the same
procedure at $T\!=\!0$ would produce.  We apply the reduction rules
recursively until the graph cannot be reduced any further.  At this
point we determine whether there exists at least one path, along which
the remaining effective couplings are nonzero, that connects the
$z\!=\!0$ and $z\!=\!L$ plane of the 3D lattice with periodic boundary
conditions in the $x$ and $y$ and open boundary conditions in the $z$
direction. The presence of such a path is a necessary, but not
sufficient condition for a SG phase.  In
Fig.~\ref{fig:percolation_at_T} we show the probability that such a
path exists as a function of $p$ for different sizes $L$.  It clearly
shows that exact cancellations also in 3D lead to $p_{\rm c}\!<\!\pSG$.

\paragraph*{Discussion.---}\hspace{-0.45cm}
An important comment is in order: optimization methods that compute
ground state energies as in Ref.~\cite{boettcher:04b} are typically
sensible to the ``unphysical'' isolated critical point at $p^*$ rather
than to the ``physical'' critical point at $\pSG$.  To detect the right
critical point these optimization methods should be modified to include
entropic effects.  In Ref.~\cite{hartmann:02f} an attempt in that
direction was made, leading to a change in physical observables. The
decimation method used in Ref.~\cite{boettcher:04b} can be easily
extended to consider first-order corrections in temperature, thus
leading to a more connected lattice after decimation.

However, even if a correct procedure is employed, taking into account
entropic effects, the presence of the spurious fixed point at $p^*$
may lead to very strong crossover effects. For example for $p \in
(\pSG,p^*)$, as long as energetic couplings are present, they dominate
the behavior of any observable; the right physical behavior, given by
entropic couplings, can be measured only at length scales where
energetic couplings have disappeared, and these scales may be
extremely large.  In the $b=3$ and $s=2$ hierarchical lattice such a
length scale is always larger than $s^{15}=32768$, roughly.
Increasing the dimensionality we find that $\pSG$ and $p^*$ get
closer, thus making the crossover length even larger.  This argument
suggests that any numerical method at zero temperature is plagued by
the unphysical fixed point up to unreachable length scales!

\begin{acknowledgments}
  We thank A.\! J.~Bray, S.~Jim\'{e}nez, I.~Kanter, F.~Krz\c{a}ka{\l}a,
  L.~Le Pera, M.~Palassini, G.~Parisi and N.~Sourlas for discussions.
  The simulations have been performed in part on CINECA's CLX cluster.
  We acknowledge support from EEC's FP6 IST contracts under IP-1935
  (EVERGROW) and IST-034952 (GENNETEC).
\end{acknowledgments}

\bibliography{refs_dEA}

\begin{thebibliography}{31}
\bibitem[{not()}]{note_gapless}
\bibinfo{note}{Models we study have an energy gap (discrete couplings), but
  entropy fluctuations make the free-energy gapless.}
\end{thebibliography}

\end{document}